\documentclass[preprint,showpacs,preprintnumbers,amsmath,amssymb]{revtex4}
\usepackage{graphicx}
\usepackage{epsfig}
\usepackage{bm}
\usepackage{amsfonts}
\usepackage{subfigure}
\usepackage{multirow}
\usepackage{float}
\usepackage{color}
\providecommand{\tabularnewline}{\\}
\begin{document}
\title{\textbf{Primordial black holes and secondary gravitational waves from generalized power-law non-canonical inflation with quartic potential}}
\author{Soma Heydari\footnote{s.heydari@uok.ac.ir} and Kayoomars Karami\footnote{kkarami@uok.ac.ir}}
\address{\small{Department of Physics, University of Kurdistan, Pasdaran Street, P.O. Box 66177-15175, Sanandaj, Iran}}
\date{\today}
%============================================Abstract===================================================
\begin{abstract}
\noindent
Here, generation of PBHs and secondary GWs from non-canonical inflation with quartic potential have been probed. It is illustrated that, quartic potential in non-canonical setup with a generalized power-law Lagrangian density can source a consistent inflationary era with the latest observational data. Besides, we show that our model satisfies the swampland criteria. At the same time, defining a peaked function of inflaton field as non-canonical mass scale parameter $M(\phi)$ of the Lagrangian, gives rise to slow down the inflaton in a while. In this span, namely Ultra-Slow-Roll (USR) stage, the amplitude of the curvature perturbations on small scales enlarges versus CMB scales. It has been illustrated that, further to the peaked aspect of the chosen non-canonical mass scale parameter, the amount of $\alpha$ parameter of the Lagrangian has enlarging impact on the amplitude of the scalar perturbations. As a consequence of adjusting three parameter Cases of this model, three Cases of PBHs in proper mass scopes to explain LIGO-VIRGO events, microlensing events in OGLE data and DM content in its totality, could be produced. In the end, power-law behavior of the current density parameter of gravitational waves $\Omega_{\rm GW_0}$  in terms of frequency has been examined. Also, the logarithmic power index as $n=3-2/\ln(f_c/f)$ in the infrared regime is obtained.
 \end{abstract}
\maketitle
%============================================Introduction===================================================
\newpage
\section{Introduction}
It is widely recognized that, quantum fluctuations of scalar field in the inflationary era
could produce primal curvature perturbations. The hastened expansion of the universe during the inflation throws the perturbed modes out of the Hubble horizon. These modes come
back to the horizon in Radiation Dominated (RD) era. The enlarged perturbed
scales can produce the ultra-dense domains in RD era. In the end, gravitational cave-in of
these domains causes to born Primordial Black Holes (PBHs) with different masses.
The earliest allusion to the context of PBHs was referred to the 1970s  \cite{zeldovich:1967,Hawking:1971,Carr:1974,carr:1975}. Thenceforth,
fortunate detection of Gravitational Waves (GWs) ensued from converging black holes with masses of  $ 30 M_\odot$ ($M_\odot$ connotes the solar mass) by LIGO-Virgo cooperation \cite{Abbott:2016-a,Abbott:2016-b,Abbott:2017-a,Abbott:2017-b,Abbott:2017-c}, has returned the PBHs notion to the center of attention. So PBHs could be considered as the origin of traced GWs. Into the bargain, the cryptic essence of Dark Matter (DM) \cite{Garrett:2011}, has inspired the researchers to ponder about PBHs as a probable candidate for the totality or a portion of DM content \cite{Heydari:2023,Wang:2023,Khlopov:2010,
OGLE-1,Kamenshchik:2019,fu:2019,Dalianis:2019,mahbub:2020,mishra:2020,Solbi-a:2021,Solbi-b:2021,Teimoori-b:2021,Laha:2019,Teimoori:2021,fu:2020,
Dalianis:2020,Pi,Garcia-Bellido:2017,Heydari:2022,Heydari-b:2022,Rezazadeh:2021,Kawaguchi:2023,Kawai:2021edk,Saridakis:2023,Clesse:2015,Kawasaki:2016,
Braglia:2020,Motohashi:2017,Sayantan-4:2023,Sayantan-6:2023,
Cai:2023,Belotsky:2014,Belotsky:2019,Ashrafzadeh:2023,Domenech:2021,Domenech-1:2020,Domenech-2:2020,
Drees:2021,Kawai:2022emp,Yuan:2021,Ahmed}.

PBHs could be produced in broad mass range. Some observational data impose constraints on their mass spectra. The localized PBHs mass spectra in   ${\cal O}(10)M_\odot$ can be restrained by LIGO-VIRGO events. The PBHs mass spectra in permissible regime of OGLE data \cite{OGLE-1,OGLE-2}, around ${\cal O}(10^{-5})M_\odot$, could be considered as sources of ultrashort-timescale microlensing events. The PBHs in the mass range ${\cal O}(10^{-16}-10^{-11})M_\odot$ could comprise the whole DM content, since there are not any observational constraint in this zone \cite{Heydari:2023,Ashrafzadeh:2023,Teimoori:2021,Teimoori-b:2021,Heydari:2022,Heydari-b:2022,fu:2019,Dalianis:2020,Solbi-a:2021,Solbi-b:2021}.

In the PBHs generation scenarios, the amplitude of the curvature perturbations power spectrum  ${\cal P}_{\cal R}$ needs to be amplified on small scales around seven order of magnitude with regard to CMB scale
\cite{Heydari:2023,OGLE-1,Kamenshchik:2019,Ashrafzadeh:2023,fu:2019,Dalianis:2019,mahbub:2020,mishra:2020,Solbi-a:2021,Solbi-b:2021,Teimoori-b:2021,Laha:2019,Teimoori:2021,
fu:2020,Dalianis:2020,Pi,Garcia-Bellido:2017,Motohashi:2017,Heydari:2022,Heydari-b:2022,
Rezazadeh:2021,Kawaguchi:2023,Kawai:2021edk,Saridakis:2023,
Clesse:2015,Braglia:2020,Kawasaki:2016,Kawai:2022emp}.
On the other side, the CMB anisotropies data of Planck 2018 \cite{akrami:2018} confine the scalar power spectrum to ${\cal P}^{*}_{\cal R}\simeq2.1 \times 10^{-9}$  at pivot scale $k_{*}=0.05~ \rm Mpc^{-1}$. So, a suitable process is needed to amplify the amplitude of the scalar power spectrum from ${\cal O}(10^{-9})$  at the CMB scale to ${\cal O}(10^{-2})$ on PBHs scales, without any conflict with observational
data. Thus far, a large numbers of techniques in the standard model of inflation \cite{Dalianis:2019,mahbub:2020,Garcia-Bellido:2017,Motohashi:2017,mishra:2020,
Rezazadeh:2021}, and beyond the standard model \cite{Heydari:2023,Ashrafzadeh:2023,Kawaguchi:2023,Kawai:2021edk,Pi,Dalianis:2020,fu:2019,Teimoori:2021,Heydari:2022,Heydari-b:2022,
fu:2020,Kamenshchik:2019,Saridakis:2023,Solbi-a:2021,Solbi-b:2021,Clesse:2015,Kawasaki:2016,
Braglia:2020,Kawai:2022emp,Teimoori-b:2021} have been offered to achieve this goal.

Secondary GWs propagate in the cosmos coeval with production of PBHs and they could
be tracked down by different GWs observatories in various frequency ranges
\cite{Heydari:2023,Kamenshchik:2019,fu:2020,Dalianis:2019,mahbub:2020,Solbi-a:2021,Solbi-b:2021,Teimoori-b:2021,Laha:2019,Teimoori:2021,fu:2020,Dalianis:2020,
Heydari:2022,Heydari-b:2022,Rezazadeh:2021,Kawaguchi:2023,Kawai:2021edk,Saridakis:2023}. Detection of secondary GWs via detectors like Square Kilometer
Array (SKA) \cite{ska}, European Pulsar Timing Array (PTA) \cite{EPTA-a,EPTA-b,EPTA-c,EPTA-d}, Laser Interferometer Space Antenna (LISA) \cite{lisa,lisa-a}  and so on, could be the indirect way to observe the PBHs. In addition, rectitude of PBHs generation models could be examined by way of the forthcoming data of these detectors.

In most of the PBHs production models, enhancement of the primordial curvature perturbations takes places in an Ultra-Slow-Roll (USR) phase of the inflationary era. In the USR phase due to a friction dominated process,  velocity of the inflaton field decreases in comparison with slow-roll regime. Usually, in this phase the slow-roll condition is broken down by the
second slow roll parameter. In this way extra time for enough growth of ${\cal P}_{\cal R}$ to produce PBHs is provided.
Basically, the USR regime in the standard model of inflation could be generated by altering the inflationary potentials. This aim could be reached either by altering the potentials to have inflection points \cite{Dalianis:2019,mahbub:2020,Garcia-Bellido:2017,Motohashi:2017}, or by setting a bump/dip for the potentials on the scales smaller than CMB scale \cite{mishra:2020,Rezazadeh:2021}. Hence, decreasing the velocity of the inflaton could take place in the vicinity of the inflection point or in the bump/dip passing. Beyond the standard model of inflation, the USR regime could be produced via applying suitable coupling functions to enhance the friction gravitationally \cite{Teimoori-b:2021,Ashrafzadeh:2023,Dalianis:2020,fu:2019,Teimoori:2021,Heydari:2022,Heydari-b:2022,fu:2020,Solbi-a:2021,Solbi-b:2021}.

The non-canonical inflationary model with power-law Lagrangian ${\cal L}(X,\phi)=X^{\alpha}-V(\phi)$, where $X$ connotes the canonical kinetic term, is a well recognized generalization for the standard model of inflation  \cite{Unnikrishnan:2012,Rezazadeh:2015,Mishra:2022,Saridakis:2023,Heydari:2023}. Therein, for  $\alpha=1$ the canonical standard model of inflation is recovered. It has been shown that, smaller obtained slow-roll parameters in this framework in comparison with canonical case can lead to the longer inflationary era and also larger scalar spectral index $(n_{s})$ and smaller tensor-to-scalar ratio $(r)$ \cite{Unnikrishnan:2012}. Thence, the steep potentials such as  quartic and exponential potentials could be resurrected in this setup \cite{Unnikrishnan:2012,Rezazadeh:2015,Mishra:2022}. On the other hand, it has been proven that in this framework allowable PBHs mass spectra and detectable GWs could be produced on small scales \cite{Saridakis:2023,Heydari:2023}.

So as to check the theoretical rectitude of the inflationary models, their compatibility with swampland criteria in addition to observational constraints have been verified. The swampland criteria emanate from string theory and comprise two conjectures namely distance conjecture and de Sitter conjecture \cite{Garg:2019,Ooguri:2019,Kehagias:2018}. The distance conjecture imposes an upper limit on the filed evolution $\Delta\phi<1$. The de Sitter conjecture imposes a lower limit on the potential gradient $|V_{,\phi}/V|>1$. The de Sitter conjecture is in conflict with the slow-roll condition  in the standard model of inflation due to $\varepsilon_1=|V_{,\phi}^2/V^2|\ll1$.

In the present work, we are interested in studying the production of PBHs and GWs in the framework of generalized power-law non-canonical inflation with quartic potential, in which the mass scale parameter of non-canonical Lagrangian $M(\phi)$ depends on the scalar field. We try to employ the field-dependent feature of $M(\phi)$ to make the model not only be consistent with observations at CMB scale but also produce PBHs on small scales.
To do this, in Sec. \ref{sec2}, fundamental relations of generalized power-law non-canonical  model are analyzed. Thence, in Sec. \ref{sec3}, observational and theoretical viability of the model thereto enhancing process of the amplitude of the scalar power spectrum at small scales  has been expounded. Thereafter, PBHs mass spectra and current density parameter spectra of secondary GWs are obtained in  Sec. \ref{sec4} and  Sec. \ref{sec5}, respectively. In the end, Sec. \ref{sec6} is allotted to the abridged results of the paper.
%==========================================non-canonical model======================================================
\section{Generalized power-law non-canonical inflation}\label{sec2}
We start by the following common action
\begin{equation}\label{action}
S=\int{\rm d}^{4}x\ \sqrt{-g}\ {\cal L}(X,\phi),
\end{equation}
in which the Lagrangian density $ {\cal L}(X,\phi)$ could be defined as varied functions of scalar field $\phi$ and kinetic term $X\equiv\frac{1}{2}\partial_{\mu}\phi\partial^{\mu}\phi$ \cite{Panotopoulos-2007,Chimento-2004,Unnikrishnan:2012,Rezazadeh:2015,Mishra:2022}.
Here, we introduce a new generalized power-law shape for the Lagrangian density as follows
\begin{equation}\label{Lagrangian}
{\cal L}(X,\phi) = X\left(\frac{X}{M(\phi)^{4}}\right)^{\alpha-1} - V(\phi),
\end{equation}
therein, the non-canonical dimensionless $\alpha$ parameter specifies deviation from canonicity.
for $\alpha=1$ the canonical version of Lagrangian (\ref{Lagrangian}) can be retrieved. The parameter $M(\phi)$ with dimension of mass, denotes the non-canonical mass scale of inflation. For the first time we consider $M(\phi)$ as a general function of scalar field instead of a constant parameter.  $V(\phi)$ is the inflationary potential.
The homogeneous and isotropic universe is described by the ensuing flat Friedmann-Robertson-Walker (FRW) metric as
\begin{equation}
\label{eq:FRW}
{\rm d}{s^2} = {\rm d}{t^2} - {a^2}(t)\left( {{\rm d}{x^2} +
{\rm d}{y^2} + {\rm d}{z^2}} \right),
\end{equation}
where $t$ is the cosmic time and  $a(t)$ denotes the scale factor. Using the flat FRW metric, the kinetic term transforms into $X ={\dot \phi ^2}/2$ (dot signifies derivative against $t$).

It is noteworthy that, the following field redefinition
\begin{equation}
\label{eq:psi}
\psi = \int M(\phi)^{\frac{2(1-\alpha)}{\alpha}}{\rm d}\phi,
\end{equation}
results in recovery of the well known form of the non-canonical Lagrangian ${\cal L}(\tilde{X},\psi) = \tilde{X}^{\alpha} - U(\psi)$ from the Lagrangian (\ref{Lagrangian}), wherein  $\tilde{X}={\dot \psi ^2}/2$ and $U(\psi)=V(\phi(\psi))$. In this way, any function $M(\phi)^4$ can be eliminated from the non-canonical kinetic term of the Lagrangian (\ref{Lagrangian}) by the field redefinition (\ref{eq:psi}). Nevertheless, some form of $M(\phi)$ in the field redefinition (\ref{eq:psi}) may cause  complexity in deriving an  analytical expression for the potential $U(\psi)$. Hence, it is reasonable to use
the Lagrangian (\ref{Lagrangian}) in such models depending on the form of $M(\phi)$.

In what follows the ensuing form of energy density $\rho _\phi$ and pressure $p_\phi$ of the scalar field could be computed from the Lagrangian (\ref{Lagrangian}) (see \cite{Unnikrishnan:2012,Rezazadeh:2015} to peruse these equations)
\begin{eqnarray}
{\rho _\phi } &=&2X\left( {\frac{{\partial {\cal L}}}{{\partial X}}} \right) - {\cal
L}= \left( {2\alpha - 1}
\right)X{\left( {\frac{X}{{{M(\phi)^4}}}} \right)^{\alpha  - 1}} +
V(\phi), \label{eq:rho}
\\
{p_\phi } &=& {\cal L}=X{\left( {\frac{X}{{{M(\phi)^4}}}} \right)^{\alpha  - 1}} - V(\phi ).\label{eq:p}
\end{eqnarray}
Utilizing Eqs. (\ref{eq:rho}) and (\ref{eq:p}), the first and second Friedmann equations are derived as
\begin{eqnarray}
\label{eq:Friedmann}
H^{2} &=&\frac{1}{3 M_{\rm p}^2}\rho _\phi= \frac{1}{3 M_{\rm p}^2}\left[\left(2\alpha-1\right)X\left(\frac{X}{M(\phi)^{4}}\right)^{\alpha-1} +\;
V(\phi)\right]~,\label{eq: FR-eqn1}\\
\dot{H} &=&\frac{-1}{2 M_{\rm p}^2}(\rho _\phi+p_{\phi}) =-\frac{1}{M_{\rm p}^2}\alpha X\left(\frac{X}{M(\phi)^4}\right)^{\alpha -1},
\label{eq: FR-eqn2}
\end{eqnarray}
in which $H\equiv \dot{a}/a $ designates the Hubble parameter and $M_{\rm p}=1/\sqrt{8\pi G}$
is the reduced Planck mass.

Using the Lagrangian (\ref{Lagrangian}) and minimizing the action (\ref{action}) with respect to scalar field $\phi$, the ensuing second order equation of motion can be derived
\begin{equation}
\label{eq:KG}
\ddot \phi  + \frac{{3H\dot \phi }}{{2\alpha  - 1}}-\left(\frac{2(\alpha-1)}{\alpha}\right)\left(\frac{M_{,\phi}}{M(\phi)}\right)\dot\phi^2
+ \left( {\frac{{V_{,\phi }}}{{\alpha (2\alpha  - 1)}}}
\right){\left( {\frac{{2{M(\phi)^4}}}{{{{\dot \phi }^2}}}} \right)^{\alpha
- 1}} = 0,
\end{equation}
wherein $(_{,\phi})$ signifies derivative against $\phi$. It is worth noting that, for $\alpha=1$, all of the preceding equations revert to their canonical versions.

In the following, the first and second  Hubble slow-roll parameters are introduced as
\begin{equation}\label{SRP}
  \varepsilon_{1} \equiv -\frac{\dot H}{H^2}, \hspace{.5cm}  \varepsilon_{2} \equiv \frac{\dot{\varepsilon_{1}}}{ H\, \varepsilon_{1}}.
\end{equation}
Under the slow-roll approximation conditions ($\{ \varepsilon_{1},\varepsilon_{2}\}\ll 1$),  kinetic energy term can be dominated by the potential energy. So, thereunder the first Friedmann equation (\ref{eq:Friedmann}) abbreviates to  \cite{Unnikrishnan:2012}
\begin{align}
\label{eq:FR1-SR}
& 3 M_{\rm p}^{2}H^2\simeq V(\phi),
\end{align}
Furthermore, under the slow-roll conditions in Eq. (\ref{eq:KG}), $\ddot \phi$ and $\dot\phi^2$ could be neglected and using Eq. (\ref{eq:FR1-SR}), the equation of motion can be recast in
\begin{align}
 \label{eq:KG-SR}
& \dot \phi  =  -
\theta {\left\{ {\left( {\frac{{{M_{\rm p}}}}{{\sqrt 3 \alpha }}}
\right)\left( {\frac{{\theta V_{,\phi }}}{{\sqrt {V(\phi )} }}}
\right){{\left( {2{M(\phi)^4}} \right)}^{\alpha  - 1}}}
\right\}^{\frac{1}{{2\alpha  - 1}}}},
\end{align}
where $\theta  = 1$ for $ V_{,\phi }>0$ and $\theta  = -1$ for
$ V_{,\phi }<0$.

Applying the definition of the first slow-roll parameter $( \varepsilon_{1} \equiv -\dot H/H^2)$ and using $dN=Hdt$ to convert the cosmic time to $e$-folding number $N$ thereto Eqs. (\ref{eq:Friedmann}), the field equation of motion (\ref{eq:KG}) takes the following form
\begin{equation}
\label{eq:KG-NC}
\phi_{,NN} + \left[\frac{3}{2\alpha -1} - \varepsilon_{1} \right]\phi_{,N} -\left(\frac{2(\alpha-1)}{\alpha}\right)\left(\frac{M_{,\phi}}{M(\phi)}\right)\phi_{,N}^2+ \frac{V_{,\phi}}{V}\left[\frac{3\alpha - (2\alpha -1)\varepsilon_{1}}{\alpha(2\alpha-1)} \right]\frac{\phi_{,N}^ {2}}{2\varepsilon_{1}}=0,
\end{equation}
wherein $(_{,N})$ and $(_{,NN})$ designate  the first and 2nd derivative against $N$, respectively.

So as to analyze the perturbations dynamics in this framework, we pursue the calculations of \cite{Garriga:1999}. Therefrom, the curvature power spectrum under the slow-roll approximation at sound horizon passing $(c_{s}k=aH)$ by comoving wavenumber $k$ in non-canonical setup is given by
\begin{equation}\label{eq:Ps-SR}
{\cal P}_{\cal R}=\frac{H^2}{8 \pi ^{2}M_{\rm p}^{2} c_{s} \varepsilon_{1}}\Big|_{c_{s}k=aH}\,,
\end{equation}
where
\begin{equation}
\label{eq:cs-NC} c_s^2 \equiv \frac{{\partial {p_\phi }/\partial
X}}{{\partial {\rho _\phi }/\partial X}} = \frac{{\partial {\cal
L}(X,\phi )/\partial X}}{{\left( {2X} \right){\partial ^2}{\cal
L}(X,\phi )/\partial {X^2} + \partial {\cal L}(X,\phi )/\partial
X}},
\end{equation}
signifies the square of the scalar perturbations sound speed  \cite{Garriga:1999,Unnikrishnan:2012}. The amplitude of the curvature power spectrum
at pivot scale $(k_{*}=0.05~\rm Mpc^{-1})$  has been constrained by Planck teamwork as
 $ {\cal P}_{\cal R}(k_{*})\simeq 2.1 \times 10^{-9}$ \citep{akrami:2018}.
For the power-law Lagrangian (\ref{Lagrangian}), the square of the sound speed (\ref{eq:cs-NC}) takes the following form
\begin{equation}
\label{eq:cs}
c_s^2 = \frac{1}{{
2\alpha  - 1 }}.
\end{equation}
It is obvious that, in order to prevent the model from classical instability we required $c_s^2>0$ which results in  $\alpha>1/2$.

Pursuant to \cite{Garriga:1999}, in this setup the scalar spectral index $n_{s}$ is obtained from the curvature power spectrum and it could be written with regard to slow-roll parameter as follows
\begin{align}\label{eq:ns-SR}
n_s-1\equiv \frac{d\ln{\cal P}_{\cal R}}{d\ln k}= -2\varepsilon_1-\varepsilon_2.
\end{align}
The scalar spectral index is observationally constrained by Planck 2018 as
$n_s= 0.9668 \pm 0.0037$ (TT,TE,EE+lowE+lensing+BK15+BAO, 68\%  CL)  \cite{akrami:2018}.

Subsequent to  \cite{Garriga:1999},  the tensor perturbations power spectrum under the slow-roll approximation at $k=aH$ in non-canonical framework is given by
\begin{equation}\label{eq:Pt-SR}
{\cal P}_{t}=\frac{2H^2}{\pi ^{2}M_{\rm p}^2}.
\end{equation}
It is evident that, the tensor power spectrum in this setup is equal to its canonical counterpart. Because The tensor power spectrum in this setup ia associated to the gravity sector of the action.
Employing the scalar  (\ref{eq:Ps-SR}) and tensor (\ref{eq:Pt-SR}) power spectra, the tensor-to-scalar ratio $r$  is calculated  as
\begin{equation}
\label{eq:r}
r\equiv\frac{{\cal P}_t}{{\cal P}_{\cal R}}= 16 c_s \varepsilon_1.
\end{equation}
There is an upper bound on the tensor-to-scalar ratio established by Planck 2018 as
$r<0.063$ (TT,TE,EE+lowE+lensing+BK15+BAO, 95\%  CL)  \cite{akrami:2018}. It is worth noting that, recently the mentioned upper bound on  $r$ is constricted by BICEP/Keck 2018
to $r<0.036$ at $95\%$ CL \cite{BK18:2021}.
%===================================enhanced Curvature Perturbation========================
\section{methodology and viability of the model }\label{sec3}
It is widely recognized that, creation of the seeds of PBHs during the inflation requires an enhancement around seven order of magnitude in the ${\cal P}_{\cal R}$ on small scales  in proportion to $ {\cal P}_{\cal R}(k_{*})$ at CMB scale. In the subsequent stage, the circumstances of occurring such enhancement in the ${\cal P}_{\cal R}$ through the non-canonical framework with power-law Lagrangian (\ref{Lagrangian}) has been expounded.
To this end, the ensuing two-parted function is chosen for the non-canonical mass scale parameter $M(\phi)$ of the generalized Lagrangian (\ref{Lagrangian})
\begin{equation}\label{eq:M}
M(\phi) = M_0\big(1-\epsilon(\phi)\big),
\end{equation}
wherein
\begin{equation}\label{eq:bump}
\epsilon(\phi) =\frac{w}{\sqrt{1+\frac{(\phi-\phi_c)^2}{b^2}}}.
\end{equation}
The $M_0$ parameter in (\ref{eq:M}) with the dimension of mass, is responsible for observational compatibility of the model on CMB scale as well as the $\alpha$ parameter. The second term of (\ref{eq:M}), $\epsilon(\phi)$ is a dimensionless peaked function of $\phi$, that is able to generate a localized peak in $\phi=\phi_{c}$ with height $\omega$ and breadth $b$. The parameters $\{\phi_{c},b\}$ have dimensions of mass and $\omega$ is dimensionless. It can be deduced that, for $\phi\neq\phi_{c}$ the $\epsilon(\phi)$ function melts away ($\epsilon(\phi)\ll 1$) and $M(\phi)$ comes back to constant $M_0$. Hence, $\epsilon(\phi)$ is responsible for slowing the inflaton down in the vicinity of $\phi=\phi_{c}$ (i.e. USR phase) and enhancing the curvature power spectrum on small scales without significant effect on CMB scale. So as to achieve this goal, we need to adjust the peak function parameter $\{\omega,b,\phi_{c}\}$. In this way, not only the compatibility of the model with observational data on CMB scale is insured, but also the USR phase could be generated to enhance the ${\cal P}_{\cal R}$ on smaller scales to produce PBHs.

It should be noted that, in \cite{Saridakis:2023} the authors have investigated production of PBH and GWs in non-canonical framework with power-law Lagrangian through the mechanism of inflection point in a class of steep-deformed
exponential potential. Two cases of PBHs are produced in the vicinity of inflection point of the potential (the USR stage) \cite{Saridakis:2023}.

In  \cite{Heydari:2023}, production of PBHs and GWs in the same framework as \cite{Saridakis:2023}, from the quartic potential in the presence of a tiny bump  has been investigated. At the moment of bum-passing three cases of PBHs were generated  \cite{Heydari:2023}. In both mentioned papers \cite{Saridakis:2023,Heydari:2023} the power-law Lagrangian contains the constant non-canonical mass scale parameter $M$, but in the present work we consider a generalized form of this framework with $M(\phi)$ as a function of $\phi$  defined by Eq. (\ref{eq:M}).

Here, we consider the quartic potential for the model
\begin{equation}\label{eq:qV}
V(\phi)=\frac{\lambda}{4}\phi^4.
\end{equation}
in which,  $\lambda\simeq 0.13$  is the dimensionless self coupling parameter  \cite{Tanabashi:2018}. The quartic potential can source chaotic inflation \cite{Linde:1983} and it has substantial reheating features \cite{Liddle:2003,Ford:1987}.
It is demonstrated that, the quartic potential in the standard model of inflation  suffers from some problems. First, because of the large self-coupling constant $\lambda\simeq0.13$,  its produced scalar fluctuations  on CMB scale cannot be compatible with the latest Planck data \cite{akrami:2018}.   On the other hand, owing to generation of the large tensor fluctuations by this potential, its predicted quantity for  $r$  cannot lie in the permitted domain of Planck  2018 data \cite{akrami:2018}. As regards
the   quartic potential cannot accommodate to feasible inflationary epoch in the standard model of inflation, in this work we attempt to revive it through the generalized power-law non-canonical setup.
Simultaneously, employing a peaked function for $M(\phi)$  in this setup, we could examine the generation of PBHs and GWs on smaller scales.

Altogether, this model is comprised of three Cases containing a collection of six parameters $\{\alpha, M_0,\lambda, \omega, \phi_c, b\}$. Regarding $\lambda\simeq 0.13$, the  $M_0$ parameter for each case could be computed via the observational constraint on the scalar power spectrum  (${\cal P}^{*}_{\cal R}\sim2.1 \times 10^{-9}$) at pivot scale $(k_{*}=0.05~\rm Mpc^{-1})$ \cite{akrami:2018}.  Three parameters $\{\omega,\phi_c,d\}$ do not have any significant effect on CMB scale. So $\alpha$ is the only free parameter which can be used to remedy the large tensor-to-scalar ratio $r$ of the model. So pursuant to \cite{Mishra:2018,Mishra:2021,Mishra:2022,Unnikrishnan:2012,Unnikrishnan:2013,Rezazadeh:2015},  we have chosen $\alpha=17$  for all Cases of our model to decrease the  $r$. It is notable that, using the chosen $\alpha$, Null Energy Condition (NEC) is preserved and the kinetic energy remains positive and dominated by potential energy throughout the inflationary era. Thus, the model is free of the ghost and tachyonic instabilities. Furthermore, the Lagrangian (\ref{Lagrangian}) satisfies the Effective Field Theory (EFT) conditions for non-canonical Lagrangian \cite{Franche:2010}. The assigned parameters for all Cases of the model
are classified in Table \ref{tab1}. Table \ref{tab2} embodies the resultant quantities for $n_s$,  $r$ and generated PBHs.
\begin{table}[H]
  \centering
  \caption{The assigned parameters for Cases A, B, and C considering the fixed $\lambda=0.13$ and $\alpha=17$.}
\begin{tabular}{ccccc}
  \hline
 $\#$ &\qquad $\omega$\qquad  & \qquad$\phi_{c}/M_{\rm p}$\qquad&\qquad$b/M_{\rm p}$\qquad&\qquad $M_{0}/M_{P}$\qquad\\[0.5ex] \hline\hline
  Case A& \qquad$0.9226$\qquad&\qquad$0.01070$ \qquad&\qquad$1.3\times10^{-6}$\qquad&\qquad $2.49\times10^{-5}$\qquad\\[0.5ex] \hline
  Case B&\qquad $0.9267$\qquad&\qquad$0.01165$\qquad &\qquad$1.2\times10^{-6}$\qquad&\qquad $2.54\times10^{-5}$\qquad   \\ \hline
  Case C&\qquad$0.9270$\qquad& \qquad$0.01236$\qquad&\qquad$1.1\times10^{-6}$\qquad& \qquad $2.53\times10^{-5}$\qquad\\ \hline
\end{tabular}
 \label{tab1}
\end{table}

\begin{table}[H]
\vspace{-0.6cm}
  \centering
  \caption{The computed quantities for Cases of Table \ref{tab1} with regard to the scalar spectral index $n_{s}$, the tensor-to-scalar ratio  $r$, the peak of the  amplitude of the scalar power spectrum ${\cal P}_{ \cal R}^\text{peak}$, the wavenumber $k_{\text{peak}} $,  the PBHs abundances $f_{\text{PBH}}^{\text{peak}}$ and masses $M_{\text{PBH}}^{\text{peak}}$. Here, the length of the inflationary era is $\Delta N=N_{\rm end}-N_{*}$ and the observable quantities $n_{s}$ and $r$ are calculated at CMB horizon passing $e$-folds number $(N_{*}=0)$.}
\begin{tabular}{cccccccc}
  \hline
$\#$ & \quad $n_{s}$\quad &\quad $r$\quad &\quad$ {\cal P}_{\cal R}^{\text{peak}}$\quad &\quad$k_{\text{peak}}/\text{Mpc}^{-1}$\quad& \quad$f_{\text{PBH}}^{\text{peak}}$\quad& \quad$M_{\text{PBH}}^{\text{peak}}/M_{\odot}$\quad&\quad$\Delta N$\\ \hline\hline
Case A &\quad0.9622\quad  &\quad0.030\quad& \quad0.037\quad & \quad$4.43\times10^{12}$ \quad&\quad 1.0000\quad &\quad$1.20\times10^{-13}$\quad&\quad 63.66\quad\\ \hline
Case B &\quad 0.9621\quad  & \quad 0.030\quad &\quad0.042\quad&\quad  $7.93\times10^{8}$ \quad&\quad0.0457\quad &\quad $3.76\times10^{-6}$ \quad&\quad 63.66\quad\\ \hline
Case C &\quad0.9621\quad  & \quad0.030\quad & \quad0.050\quad & \quad$4.46\times10^{5}$\quad &\quad 0.0013\quad &\quad$11.87$ \quad&\quad 63.71\quad\\ \hline
\end{tabular}
\label{tab2}
\end{table}
\noindent
To the best of our knowledge, deficiencies of the Hot Big Bang (HBB) theory could be rectified through a feasible inflationary epoch with a length of 60-70 $e$-folds number \cite{Guth:1981,Liddle:2003,Linde:1983}.  The last column of Table \ref{tab2} embodies the length of inflation ($\Delta N=N_{\rm end}-N_{*}$) apropos to Cases of Table \ref{tab1}, from the CMB horizon passing point ($N_{*}=0$) to the end of inflation ($N_{\text{end}}$). Whenever  the first slow-roll parameter meets one ($\varepsilon_{1}=1$), the inflationary era ends for Cases of this model (see the $\varepsilon_{1}$ plot in Fig. \ref{fig:e1,e2,H}).

Concerning the background evolution, the second Friedmann equation (\ref{eq: FR-eqn2}) and  the equation of motion (\ref{eq:KG-NC}) should be solved concurrently, via substituting the inflationary potential (\ref{eq:qV}) and the defined $M(\phi)$ from (\ref{eq:M})-(\ref{eq:bump}). The required initial conditions are acquired from the slow-roll equations (\ref{eq:FR1-SR})-(\ref{eq:KG-SR}) with the potential (\ref{eq:qV}).
Thereafter, evolutions of scalar field $\phi$, its derivative $\phi_{,N}$, the first and second slow-roll parameters ($\varepsilon_1$, $\varepsilon_2$) against the $e$-folds number have been plotted in Fig. \ref{fig:e1,e2,H} for Cases A (red lines), B (green lines) and C (blue lines).
A small USR region in the proximity of $\phi=\phi_{c}$ can be seen in each graph of Fig \ref{fig:e1,e2,H}. The short-living USR stage lasts for about two $e$-folds numbers. In this region, the inflaton speed is decreased and goes toward zero (see $\phi_{,N}$ in Fig. \ref{fig:e1,e2,H}). Hence, an approximately  flat region in the graph of the field evolution is produced for each Case of the model (notice $\phi$ in Fig. \ref{fig:e1,e2,H}). The inflationary era of our model is comprised of three consecutive phases namely first SR phase, intermediate USR phase and final SR phase. Following \cite{Cai:2018}, the sharpness of transitions between these phases can be inferable from the evolution of $\varepsilon_2$. As it can be seen from the $\varepsilon_2$ graph of Fig \ref{fig:e1,e2,H}, in the beginning of the USR stage $\varepsilon_2$ takes the negative values then it tends to positive values larger than one, at last it returns to zero at the beginning of the final SR phase. Hence, according to \cite{Cai:2018} we conclude that, the sharp transitions take place between three phases of the inflationary era.
As regards the reduction in the $\varepsilon_1$ during the USR phase (see $\varepsilon_1$ in Fig. \ref{fig:e1,e2,H}), according to Eq. (\ref{eq:Ps-SR}) the amplification of ${\cal P}_{ \cal R}$  can be presumable in this phase. Into the bargain, the value of $\varepsilon_{2}$  for each Case of the model oversteps its bound (slow-roll condition $\varepsilon_{2}\ll1$ ) in USR region (see $\varepsilon_2$ in Fig. \ref{fig:e1,e2,H}). Ergo, in the USR regime, the slow-roll conditions is violated by $\varepsilon_2$ even though it is held by  $\varepsilon_1$. Albeit, the validity of the slow-roll conditions in the vicinity of CMB scale ($N_{*}=0$), is obvious from the  $\varepsilon_1$ and  $\varepsilon_2$ graphs of Fig.
\ref{fig:e1,e2,H}.
\begin{figure*}
\begin{minipage}[b]{1\textwidth}
\vspace{-1cm}
\subfigure{ \includegraphics[width=.487\textwidth]%
{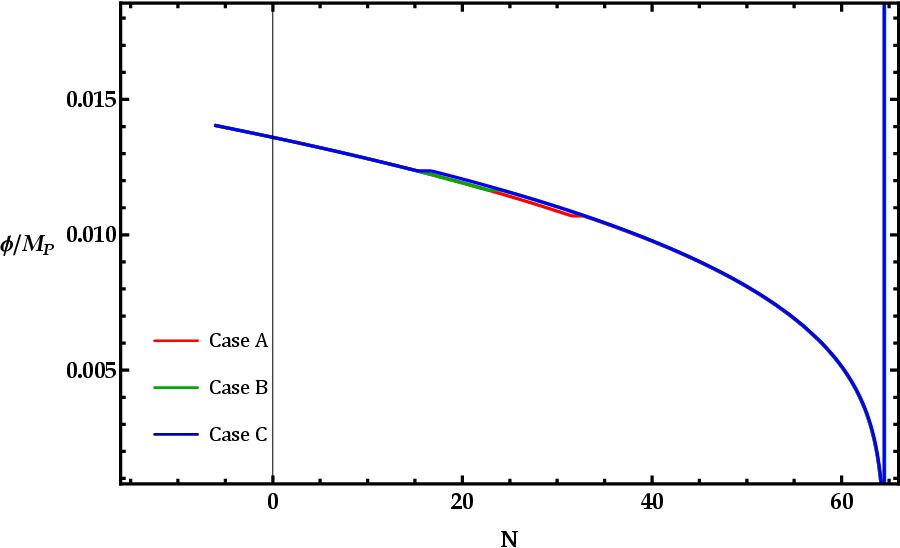}}\hspace{.1cm}
\subfigure{ \includegraphics[width=.482\textwidth]%
{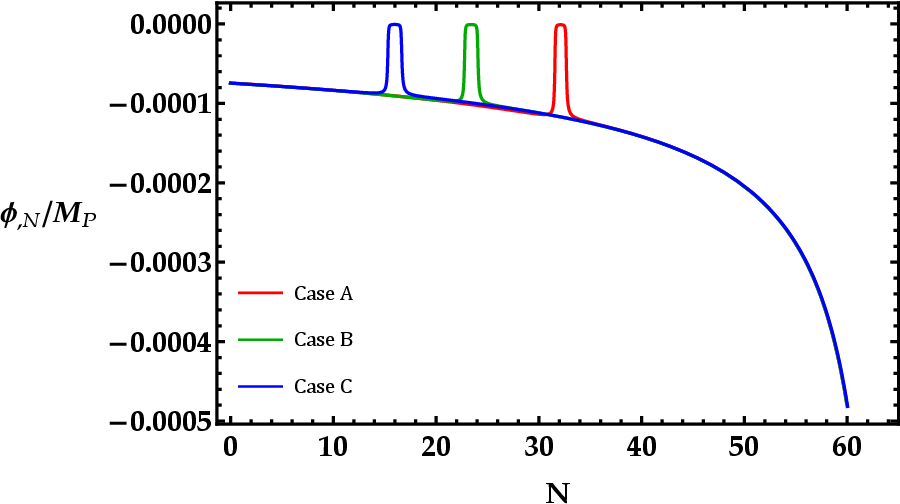}}\hspace{.1cm}
\subfigure{\includegraphics[width=.485\textwidth]%
{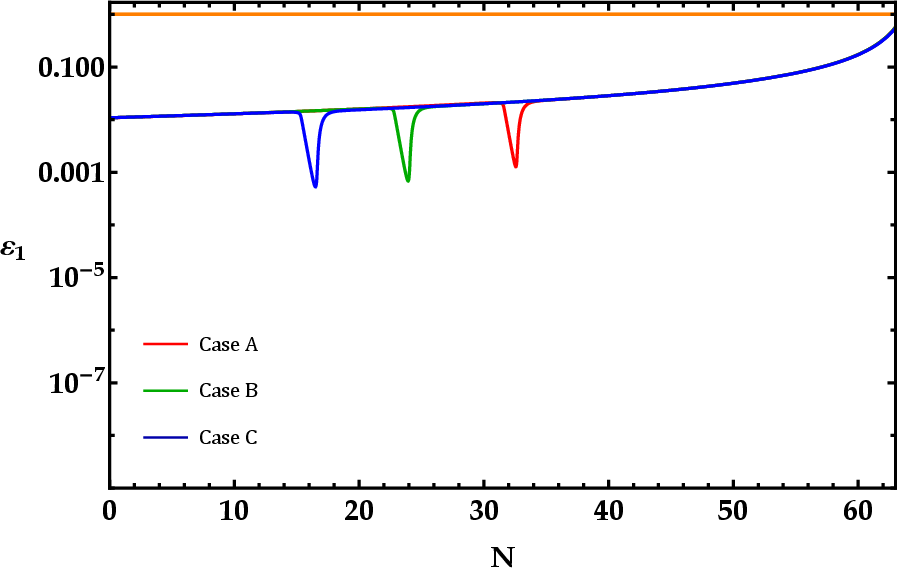}} \hspace{.1cm}
\subfigure{ \includegraphics[width=.485\textwidth]%
{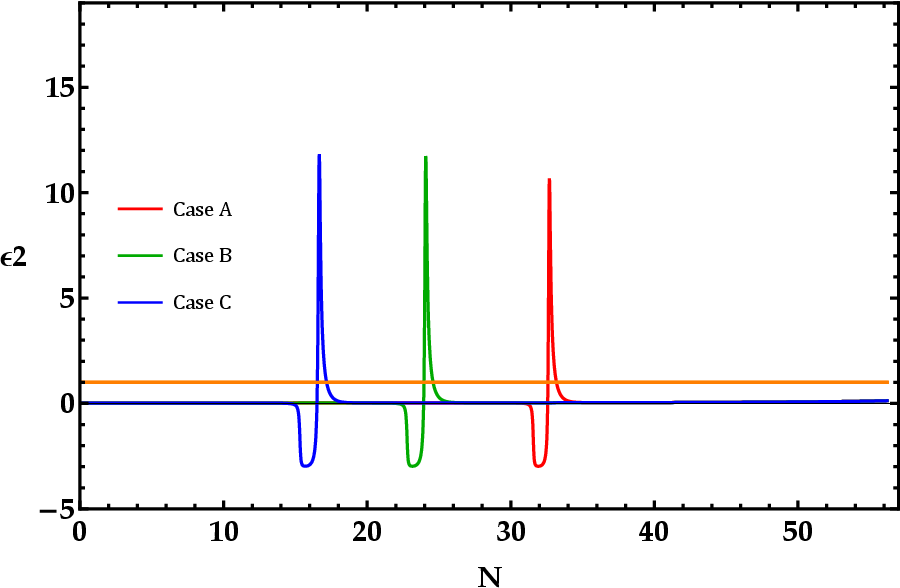}}\hspace{.1cm}
\end{minipage}
\caption{Evolution of the scalar field $\phi$ and its derivative $\phi_{,N}$ as well as the first $\varepsilon_1$ and second $\varepsilon_2$ slow-roll parameters in regard of the $e$-folds number $N$ for the Cases A (red lines), B (green lines) and C (blue lines).}
\label{fig:e1,e2,H}
\end{figure*}
Consequently, it is allowable to compute the scalar spectral index $n_s$ and tensor-to-scalar ratio $r$, applying Eqs. (\ref{eq:ns-SR}) and (\ref{eq:r}) under the slow-roll approximation for the model.
It is inferable from the listed quantities in Table \ref{tab2} that,  $n_{s}$ and $r$ apropos of each Case of the model can be compatible with Planck 2018 data (TT,TE,EE+lowE+lensing+BK15+BAO, 95\%  CL) \cite{akrami:2018}. Furthermore, the values of $r$ apropos of all Cases are compatible with BICEP/Keck 2018 data ($r<0.036$ at 95$\%$ CL) \cite{BK18:2021}. Hence, the  quartic potential through the generalized power-law non-canonical setup is resurrected in light of the latest observational data.

\begin{figure*}
\begin{minipage}[b]{1\textwidth}
\subfigure{ \includegraphics[width=.48\textwidth]%
{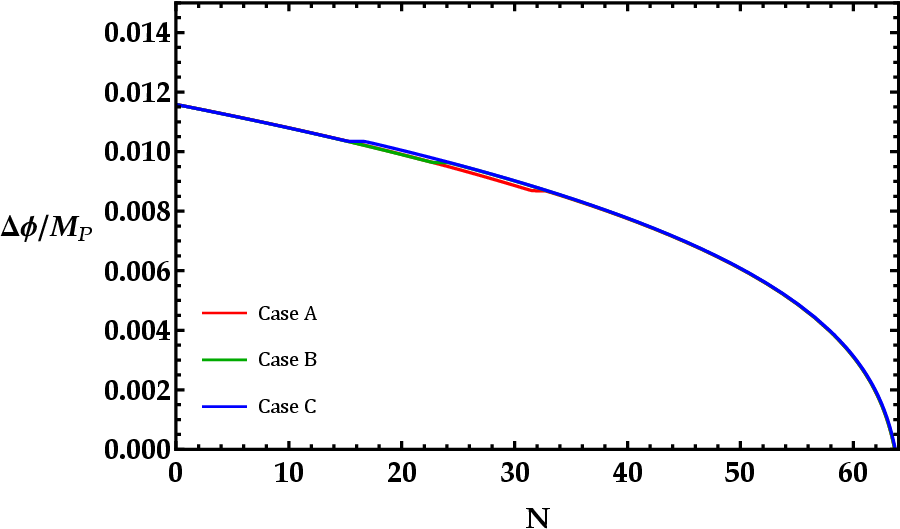}}\hspace{.1cm}
\subfigure{ \includegraphics[width=.48\textwidth]%
{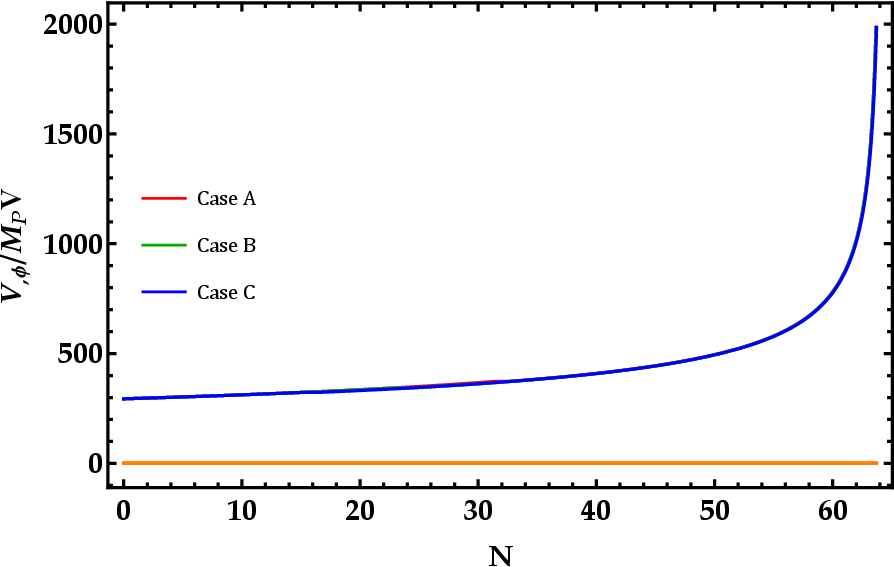}}\hspace{.1cm}
\end{minipage}
\caption{Legitimacy of swampland criteria for all Cases of the model (see plots legends), consist of distance conjecture $\Delta\phi<1$ (left panel) and de Sitter conjecture $|V_{,\phi}/V|>1$ (right panel).}
\label{fig:s}
\end{figure*}

In this stage, by considering the swampland criteria, theoretical viability of the model in addition to observational consistency can be checked. The swampland criteria originate from string theory and are comprised of two theoretical conjectures so-called distance conjecture and de Sitter conjecture  \cite{Garg:2019,Ooguri:2019,Kehagias:2018}. The distance conjecture leads to an upper bound on scalar field evolution during the inflationary era as $\Delta\phi<1$.
The de Sitter conjecture bounds the potential gradient as $|V_{,\phi}/V|>1$. Employing the field evolution, $\Delta\phi=\phi(N)-\phi(N_{\rm end})$ is calculated and plotted in Fig. \ref{fig:s} during the inflationary era. It is obvious from this plot that, the distance conjecture ($\Delta\phi<1$) is held throughout the inflationary era for all Cases and the field distance $\Delta\phi$ decrease to the end of inflation. As for the second conjecture, the potential gradient is plotted in Fig. \ref{fig:s} as well. It is inferred from this plot that, the de Sitter conjecture ($|V_{,\phi}/V|>1$) is established all over the inflationary era for all Cases and the potential gradient $|V_{,\phi}/V|$ increases from the CMB passing $e$-fold number to the end of inflation.

As described erstwhile, in the USR domain, validity of the slow-roll conditions is contravened by $\varepsilon_2$. Ergo, it is unallowable to compute the curvature power spectrum via Eq. (\ref{eq:Ps-SR}) governed by the slow-roll conditions. Thus, the evolution of the curvature perturbations should be evaluated through the subsequent Mukhanonv-Sasaki (MS) equation \cite{Garriga:1999} all over the inflationary era
\begin{equation}\label{eq:MS}
 \upsilon^{\prime\prime}_{k}+\left(c_{s}^2 k^2-\frac{z^{\prime\prime}}{z}\right)\upsilon_k=0,
\end{equation}
therein the prime connotes derivative against the conformal time $\eta\equiv\int {a^{-1}dt}$, and
\begin{equation}\label{eq:z}
 \upsilon_k\equiv z {\cal R}_k, \hspace{1cm} z \equiv \frac{a\,\left(\rho_{_{\phi}}+ p_{_{\phi}}\right)^{1/2}}{c_{_s}H}.
\end{equation}
In order to solve the MS equation (\ref{eq:MS}) numerically, it is necessary to take the ensuing Bunch-Davies vacuum state \cite{Garrett:2011} into account as the inceptive condition on the scales deep interior the horizon
\begin{equation}\label{eq:Bunch}
\upsilon_k\simeq\frac{e^{-i c_{s}k\eta}}{\sqrt{2c_s k}}, \;\;  (aH\ll c_sk).
\end{equation}
Thenceforth, the curvature perturbations power spectrum is calculated from the subsequent equation
\begin{equation}\label{eq:PsMS}
{\cal P}_{\cal R}\equiv\frac{k^3}{2\pi^2}\big|{\cal R}_k^2\big|=\frac{k^3}{2\pi^2}\left|{\frac{{\upsilon_k}^2}{z^2}}\right|.
\end{equation}
\begin{figure*}
\centering
\scalebox{0.6}[0.6]{\includegraphics{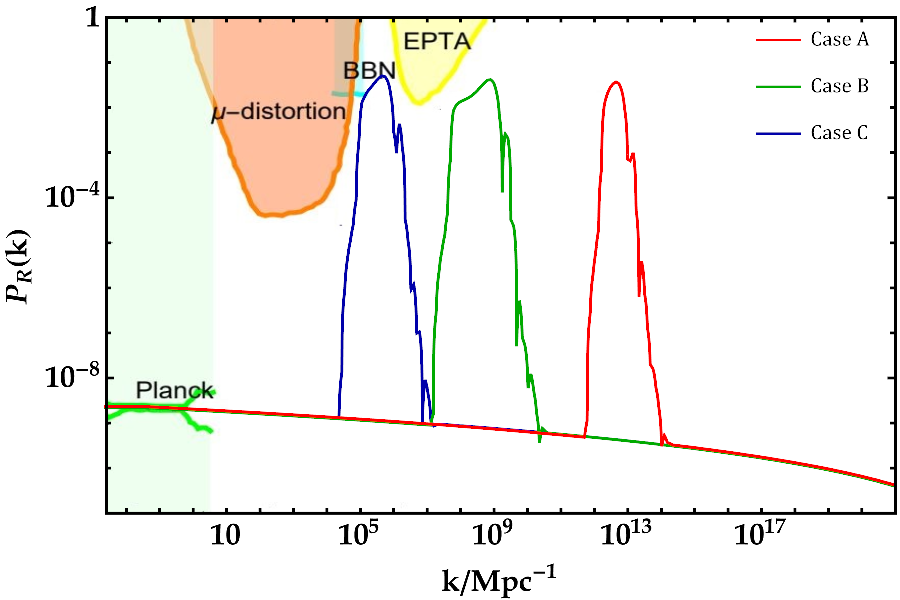}}
\caption{The curvature perturbations power spectra computed through the numerical solutions of the MS equation (\ref{eq:MS}) against the comoving wavenumber $k$ respecting the Cases A (red line), B (green line) and C (blue line).  The light-green, yellow, cyan, and orange sectors are excluded by the CMB constraints \cite{akrami:2018},  PTA constraints \cite{Inomata:2019-a},   the impact on neutron-to-proton ration during the Big Bang Nucleosynthesis (BBN) \cite{Inomata:2016,Nakama:2014,Jeong:2014}, and the $\mu$-distortion of CMB \cite{Fixsen:1996,Chluba:2012}, respectively.}
\label{fig:ps}
\end{figure*}
The acquired quantities for  the curvature power spectra  ${\cal P}_{\cal R}^{\rm peak}$ and relevant comoving wavenumbers $k_{\rm peak}$ at the peak location (matching with $\phi=\phi_c$) for each  Case of the model have been classified in Table \ref{tab2}.
Subsequently, the attained curvature power spectra ${\cal P}_{\cal R}$ with regard to comoving wavenumber  $k$ thereto the observational confinements have been plotted in  Fig. \ref{fig:ps} respecting the Cases A (red line), B (green line) and C (blue line).
It can be inferable form this figure that, the curvature power spectra respecting all Cases of the model in the environs of the CMB scale $k\sim0.05~ \rm Mpc^{-1}$, are consistent with Planck restriction  $ {\cal P}_{\cal R}(k_{*})\simeq 2.1 \times 10^{-9}$ \citep{akrami:2018}.
Moreover, the amplitudes of ${\cal P}_{\cal R}$ for all Cases during the USR domain in the vicinity of $\phi=\phi_c$ , undergo  increases to ${\cal O}(10^{-2})$ which are sufficient to produce PBHs in different masses.

\begin{figure}[H]
\centering
%\vspace{-0.9cm}
\scalebox{0.6}[0.6]{\includegraphics{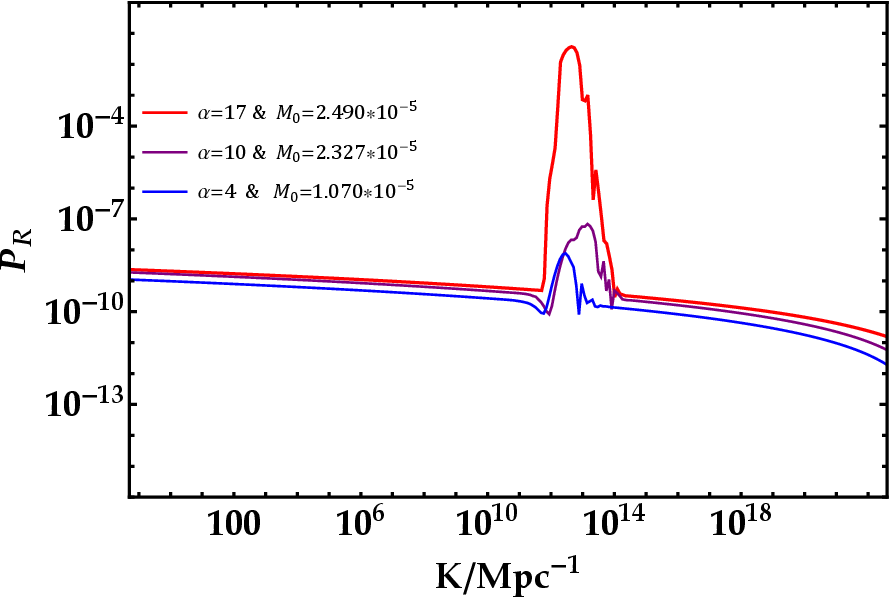}}
%\vspace{-0.6cm}
\caption{The relation between  ${\cal P}_{\cal R}$ and non-canonical $\alpha$ parameter as to  Case A of Table \ref{tab1}. The amplitude of ${\cal P}_{\cal R}$ grows up for greater $\alpha$ in USR domain (see the different values of $\alpha$ in the legend of the plot).}
\label{fig:ps_a}
\end{figure}

It should be mentioned that, the $\alpha$ parameter pertinent to the Lagrangian (\ref{Lagrangian}) could have the amplifying impacts on the amplitude of the curvature power spectra in USR domain, with no significant interference in the CMB scale. In Fig. \ref{fig:ps_a}, the relation between the amplitude of ${\cal P}_{\cal R}$ and the $\alpha$ parameter for Case A,  has been portrayed. It is deduced from this figure that, the larger $\alpha$ yields the greater ${\cal P}_{\cal R}$ in the USR domain, without any serious disturbance on CMB scale.
%===================================PBH Mass spectra==================================================
\section{PBHs mass spectra}\label{sec4}
The current section is devoted to evaluate the mass spectra of produced PBHs through the  quartic potential in non-canonical framework with the generalized power-law Lagrangian (\ref{Lagrangian}).
As described in the preceding sections, the seeds of PBHs generate in the inflationary era from the enhanced scalar perturbations. The perturbed scales leave the horizon during the inflation and thereafter when they revert to the horizon in RD era, the ultra-dense districts could be produced. Eventually, PBHs can be born from the gravitational cave-in of the mentioned districts. The mass of PBHs could be defined as a portion of the horizon mass via the subsequent equation
\begin{align}\label{Mpbheq}
M_{\rm PBH}(k)=\gamma\frac{4\pi}{H}\Big|_{c_{s}k=aH} \simeq M_{\odot} \left(\frac{\gamma}{0.2} \right) \left(\frac{10.75}{g_{*}} \right)^{\frac{1}{6}} \left(\frac{k}{1.9\times 10^{6}\rm Mpc^{-1}} \right)^{-2},
\end{align}
therein $\gamma=\big(\frac{1}{\sqrt{3}}\big)^{3}$ connotes the collapse efficiency \cite{carr:1975} and  $g_{*}=106.75$ implies the effective number of relativistic degrees of freedom.
The fractional abundance $f_{\rm{PBH}}(M)$ of PBHs  specifies that what percentage of the cosmos DM content could be allotted to PBHs and can be obtained from the ensuing equation
\begin{equation}\label{fPBH}
f_{\rm{PBH}}(M)\simeq \frac{\Omega_{\rm {PBH}}}{\Omega_{\rm{DM}}}= \frac{\beta(M)}{1.84\times10^{-8}}\left(\frac{\gamma}{0.2}\right)^{3/2}\left(\frac{g_*}{10.75}\right)^{-1/4}
\left(\frac{0.12}{\Omega_{\rm{DM}}h^2}\right)
\left(\frac{M}{M_{\odot}}\right)^{-1/2},
\end{equation}
wherein $\Omega_{\rm {DM}}h^2\simeq0.12$ implies the defined current DM density parameter by Planck data \cite{akrami:2018}. Furthermore, $ \beta(M)$ indicates the generation rate of PBHs and it can be calculated from the Press-Schechter formalism for gaussian distribution of the primal perturbations as follows
 \cite{Tada:2019,young:2014}
\begin{equation}\label{betta}
  \beta(M)=\int_{\delta_{c}}\frac{{\rm d}\delta}{\sqrt{2\pi\sigma^{2}(M)}}e^{-\frac{\delta^{2}}{2\sigma^{2}(M)}}=\frac{1}{2}~ {\rm erfc}\left(\frac{\delta_{c}}{\sqrt{2\sigma^{2}(M)}}\right).
\end{equation}
Here "erfc" is the error function complementary and $\delta_{c}=0.4$ indicates the density threshold  \cite{Musco:2013,Harada:2013}. Moreover,  $\sigma^{2}(M)$ denotes the coarse-grained density contrast smoothed on the scale $k$ as
\begin{equation}\label{sigma}
\sigma_{k}^{2}=\left(\frac{4}{9} \right)^{2} \int \frac{{\rm d}q}{q} W^{2}(q/k)(q/k)^{4} {\cal P}_{\cal R}(q),
\end{equation}
in which ${\cal P}_{\cal R}$ is the curvature power spectrum and   $W(x)=\exp{\left(-x^{2}/2 \right)}$ denotes the Gaussian window.

\begin{figure}[H]
\centering
\includegraphics[scale=0.6]{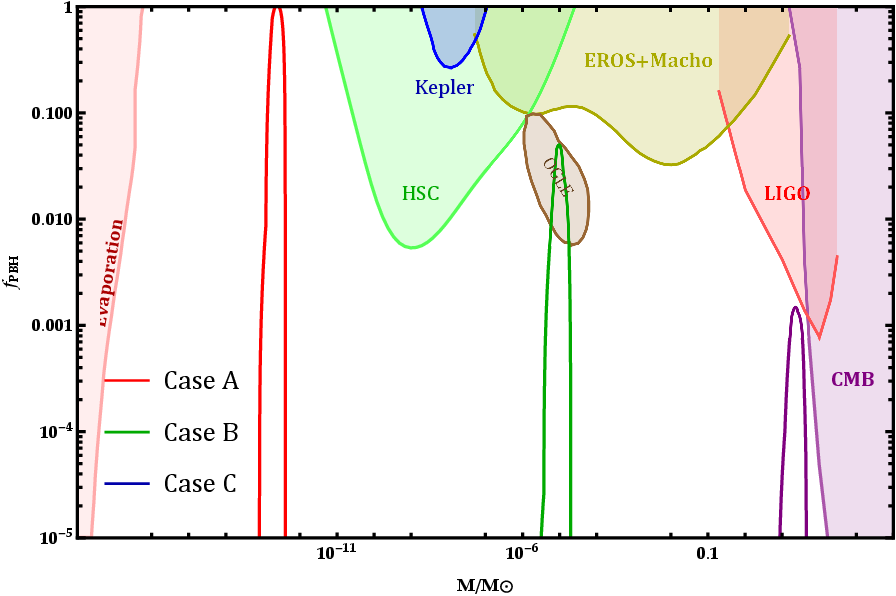}
\caption{The PBHs mass spectra relevant to Cases A (red line), B (green line) and C (blue line). The colory districts demonstrate the present observational restriction on the PBHs abundance. Some of these districts are excluded by observations of CMB \cite{CMB} (purple district), LIGO-VIRGO event \cite{Abbott:2019,Chen:2020,Boehm:2021,Kavanagh:2018} (red district), microlensing events via MACHO \cite{MACHO}, EROS \cite{EORS}, Kepler \cite{Kepler}), Icarus \cite{Icarus}, OGLE \cite{OGLE-1,OGLE-2}, Subaru-HSC \cite{subaro} (green district), and PBHs evaporation \cite{EGG, Laha:2019,Clark,Shikhar:2022,Dasgupta} (pink district). The only allowable district is the ultrashort-timescale microlensing events in the OGLE data \cite{OGLE-1,OGLE-2} (brown district).}
\label{fig-fpbh}
\end{figure}

In the next, the PBHs abundance for each Case of Table \ref{tab1} can be obtained from the  Eqs. (\ref{Mpbheq})-(\ref{sigma}), by substituting the attained ${\cal P}_{\cal R}$ from the MS equation (\ref{eq:MS}) in Eq. (\ref{sigma}). The consequent quantities for $f_{\rm PBH}$ and $M_{\rm PBH}$ (pertinent to peak location) have been classified in Table \ref{tab2} for each Case of Table \ref{tab1}.  Additionally, the PBHs mass spectra thereto the observational domains have been schemed in Fig. \ref{fig-fpbh} for all Cases of the model. As it can be seen from Table \ref{tab2}, there is a huge difference in the PBH
masses of the cases A, B and C. It can be inferred from Eq. (\ref{Mpbheq}) that, PBHs masses are proportional to
the scales $k$ wherein the PBHs are produced. Originally these scales are pertinent to the peak scales of the scalar power spectra. So we got the huge differences in the PBHs masses, because of the differences between the wavenumbers of the peaks of the scalar power spectra. The peak position of the scalar power spectrum defines the mass of produced PBH. The key parameter which is responsible to specify the peak position of scalar power spectrum is $\phi_c$ for each case of the model. So the parameter $\phi_c$ plays a key role in specifying the mass of produced PBH.

It can be inferable from Fig. \ref{fig-fpbh} that, the consequent PBHs mass spectrum from the parameter Case A could be a promising claimant for $99\%$ of DM content.
The anticipated PBHs mass spectrum for the parameter Case B has placed in the permissible zone of the OGLE data \cite{OGLE-1,OGLE-2}, and it could be proper to be considered as the origin of ultrashort-timescale microlensing events. The foretold PBHs mass spectrum concerning the parameter Case C, has situated in the sensitivity zone of LIGO-VIRGO and its concurrent GWs would be traced by these observatories.

%===================================secondary Gravitational waves=========================================
\section{ secondary gravitational waves}\label{sec5}
Propagated secondary GWs in the cosmos, could be the further upshot of reentry of the perturbed modes to the horizon in the RD era. Whereas, the concurrent GWs with PBHs production could be tracked down by multifarious detectors, so they can be considered as an indirect way to detect PBHs. The present section is earmarked for scrutinizing the secondary GWs coeval with PBHs production in the generalized power-law non-canonical model with the   quartic potential.

It can be shown that the present energy density of the secondary GWs is given by \cite{Kohri:2018}
\begin{eqnarray}\label{OGW}
&\Omega_{\rm{GW}}(\eta_c,k) = \frac{1}{12} {\displaystyle \int^\infty_0 dv \int^{|1+v|}_{|1-v|}du } \left( \frac{4v^2-(1+v^2-u^2)^2}{4uv}\right)^2\mathcal{P}_{\cal R}(ku)\mathcal{P}_{\cal R}(kv)\left( \frac{3}{4u^3v^3}\right)^2 (u^2+v^2-3)^2\nonumber\\
&\times \left\{\left[-4uv+(u^2+v^2-3) \ln\left| \frac{3-(u+v)^2}{3-(u-v)^2}\right| \right]^2  + \pi^2(u^2+v^2-3)^2\Theta(v+u-\sqrt{3})\right\}\;,
\end{eqnarray}
therein  $\Theta$ indicates the Heaviside theta function and $\eta_{c}$ implies the  termination time of the growth of  $\Omega_{\rm{GW}}$.
The present GWs energy spectrum and its counterpart at $\eta_{c}$
are associated together through  \cite{Inomata:2019-a}
\begin{eqnarray}\label{OGW0}
\Omega_{\rm GW_0}h^2 = 0.83\left( \frac{g_{*}}{10.75} \right)^{-1/3}\Omega_{\rm r_0}h^2\Omega_{\rm{GW}}(\eta_c,k)\;,
\end{eqnarray}
in which $\Omega_{\rm r_0}h^2\simeq 4.2\times 10^{-5}$ indicates the present radiation density parameter and $g_{*}\simeq106.75$ specifies the effective degrees of freedom in the energy density at $\eta_c$. The relation between frequency and wavenumber is
\begin{eqnarray}\label{k_to_f}
f=1.546 \times 10^{-15}\left( \frac{k}{{\rm Mpc}^{-1}}\right){\rm Hz}.
\end{eqnarray}

Thenceforth, the present density parameter spectra of the secondary gravitational waves  $\Omega_{\rm GW_0}$ concurrent with PBHs respecting the all Cases of Table \ref{tab1} could be attained, through the Eqs. (\ref{OGW})-(\ref{k_to_f}) and the obtained  ${\cal P}_{\cal R}$  from the  MS equation (\ref{eq:MS}).

\begin{figure}[H]
\centering
\includegraphics[scale=0.5]{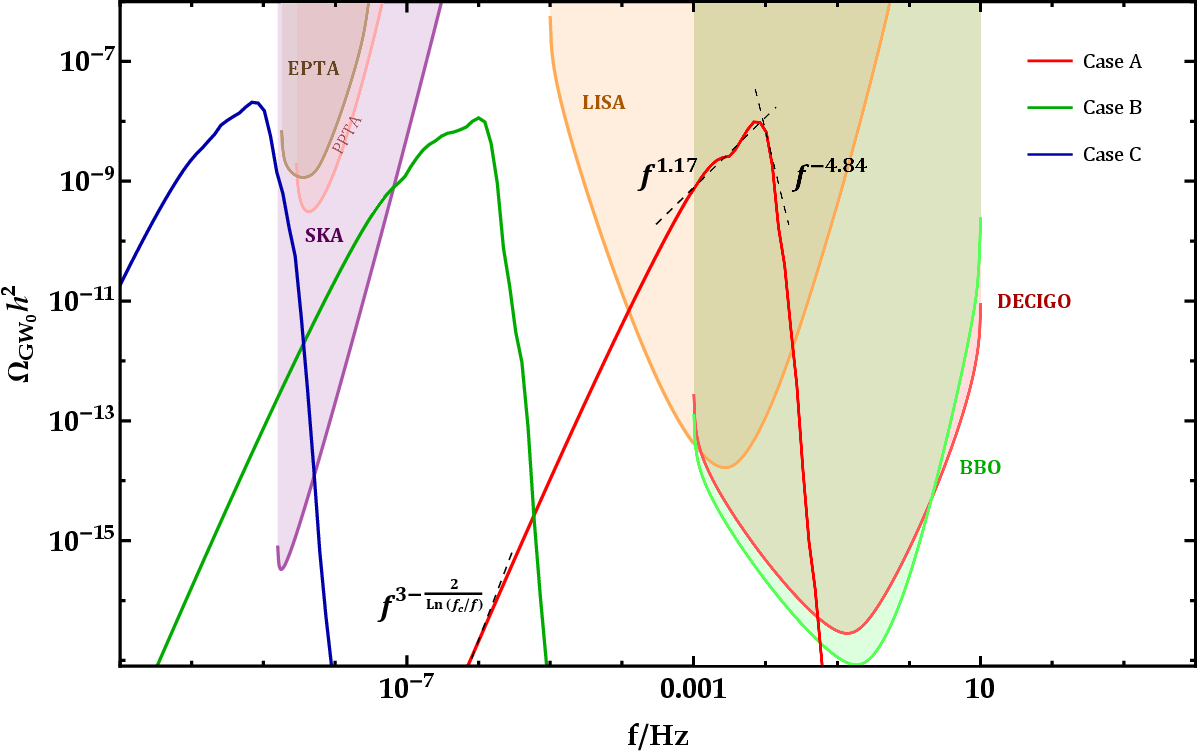}
\vspace{-0.5em}
\caption{ The acquired present energy density spectra of secondary gravitational waves  $\Omega_{\rm GW_0}h^2$ against the frequency respecting the parameter Cases A (red line), B (green line) and C (blue line) of Table \ref{tab1}, in addition to the observational regions of the GWs detectors such as EPTA (brown domain), SKA (purple domain), LISA (orange domain), DECIGO (red domain) and BBO (green domain).  The  power-law incline of the $\Omega_{\rm GW_0}h^2$ spectrum has been schemed via dashed black lines in three frequency regions for Case A.}
\label{fig-omega}
\end{figure}

The consequent  $\Omega_{\rm GW_0}$ spectra have been plotted in  Fig. \ref{fig-omega} for Cases A (red line), B (green line), C (blue line), thereto sensitivity domains of GWs detectors like SKA (purple domain) \cite{ska}, EPTA (brown domain) \cite{EPTA-a,EPTA-b,EPTA-c,EPTA-d}, LISA (orange domain) \cite{lisa,lisa-a},  BBO (green domain) \cite{Yagi:2011,BBO:2003} and DECIGO (red domain) \cite{Yagi:2011,Seto:2001}. The predicted $\Omega_{\rm GW_0}$ spectrum of parameter Case A, has lied in the sensitivity domain of LISA. Moreover, the resultant $\Omega_{\rm GW_0}$ spectra for Cases B and C have situated in the sensitivity domain of  SKA detector (see Fig. \ref{fig-omega}). Thus, the rectitude of this model could be verified in light of approaching observations of these detectors.

In the last stage, the slopes of $\Omega_{\rm GW_0}$ spectra at different frequency regions have been estimated for all Cases of the model. It has been exhibited that, the inclination of $\Omega_{\rm GW_0}$ spectrum could be matched with a power-law function of frequency as  $\Omega_{\rm GW_0} (f) \sim f^{n} $ \cite{fu:2020,Xu,Kuroyanagi}.
The estimated quantities for the power indexes $n$ for all Cases of the model in three frequency regions like $f\ll f_{c}$,  $f<f_{c}$ and  $f>f_{c}$ have been arranged in
Table \ref{table:GWs} ($f_{c}$ denotes the frequency of the peak of $\Omega_{\rm GW_0}$ spectrum). It can be inferred that, the consequent quantities for $n$ in the infrared region $f\ll f_{c}$ could be compatible with the logarithmic equation $n=3-2/\ln(f_c/f)$  acquired in \cite{Yuan:2020,shipi:2020,Yuan:2023}.
\begin{table}[H]
  \centering
  \caption{The consequent values for frequencies and heights of the peaks of $\Omega_{\rm GW_0}h^2$ spectra, in addition to the power index $n$ in frequency regions $f\ll f_{c}$,  $f<f_{c}$ and  $f>f_{c}$ respecting Cases A, B and C.}
\scalebox{1}[1] {
\begin{tabular}{cccccc}
\hline
\#  & $\qquad\qquad$ $f_{c}/{\rm Hz}$ $\qquad\qquad$ & $\quad$ $\Omega_{\rm GW_0}h^2\left(f_{c}\right)$ $\quad$ & $\quad$ $n_{f\ll f_{c}}$ $\quad$ & $\quad$ $n_{f<f_{c}}$ $\quad$ & $\quad$ $n_{f>f_{c}}$\tabularnewline
\hline
\hline
Case A & $6.86\times10^{-3}$ & $9.73\times10^{-9}$ & $3.00$ & $1.17$ & $-4.84$ \tabularnewline
\hline
Case B & $1.00\times10^{-6}$ & $1.14\times10^{-8}$ & $2.94$ & $1.10$ & $-1.49$\tabularnewline
\hline
Case C & $6.91\times10^{-10}$ & $2.07\times10^{-8}$ & $2.85$ & $1.07$ & $-1.45$\tabularnewline
\hline
\end{tabular}
    }
  \label{table:GWs}
\end{table}

As it can be seen from the listed results in Table \ref{table:GWs}, there are large differences between the frequencies of GWs. The secondary GWs are produced concurrent with PBHs generation. The frequencies of GWs are relevant to the wavenubers $k$ through Eq. (\ref{k_to_f}). As we mentioned previously, the key parameter to specify the scale $k$ for each case is $\phi_c$. On the other word the frequencies of the peaks of the GWs spectra are pertinent to peaks positions of scalar power spectra, which are specified by $\phi_c$.
%========================================Conclusions=====================================================
\section{Conclusions}\label{sec6}
This study is devoted to exhibit creation of PBHs from the  quartic potential in non-canonical inflationary model with a generalized power-law lagrangian (\ref{Lagrangian}). The observational anticipations of the  quartic potential on CMB scale, through the power-law non-canonical setup have been remedied.
The consequent quantities for scalar spectral index $n_s$ and tensor-to-scalar ratio $r$ for all Cases of the model could place in the allowable domain defined by Planck 2018 (TT,TE,EE+lowE+lensing+BK15+BAO, 95\%  CL) \cite{akrami:2018}.
Additionally, the $r$ quantities respecting all Cases of the model could be compatible with  the latest limitation  $r<0.036$ of BICEP/Keck 2018 data at 95$\%$ CL \cite{BK18:2021} (notice Table \ref{tab2}).

Choosing the non-canonical mass scale parameter $M(\phi)$ of the Lagrangian (\ref{Lagrangian}) as a two-parted peaked function of $\phi$  (\ref{eq:M})-(\ref{eq:bump}) leads to make the inflaton slow down in an USR region (in the vicinity of the peak position $\phi=\phi_c$) with no effect on CMB scale. Thus, after adjusting the parameters of the model according to Table \ref{tab1}, the sufficient enhancement in the scalar power spectrum during the USR region could be produced to born PBHs seeds. In this way not only the quartic potential could originate a feasible inflationary era, but also the PBHs and GWs seeds could be generated.
Note that our model consists of  $6$ free parameters $\{\alpha, M_0,\lambda, \omega, \phi_c, b\}$. Since $\lambda=0.13$ is fixed, the parameter $M_0$ can be obtained from the CMB normalization constraint. So $\alpha$ is the only free parameter which can be fine tuned to remedy the results of $n_s$ and $r$ for the quartic potential. Three parameters $\{\omega,\phi_c,d\}$  of the peak function do not have any significant effect on CMB scale. These three
parameters should be fine tuned to produce PBHs in different mass range with
allowable abundances. The $\omega$ parameter is responsible for the height of the peak of ${\cal P}_{\cal R}$ and subsequent PBH abundance, The $\phi_c$ parameter is defined the peak position of the ${\cal P}_{\cal R}$ and the mass of produced PBH and $d$ parameter is responsible for duration of USR stage and breadth of the peak of the ${\cal P}_{\cal R}$.
The USR stage in this model lasts for about two $e$-folds numbers. Following \cite{Cai:2018}, from the behavior of $\varepsilon_2$ we inferred that, the sharp transitions could occur between three phases of the inflationary era of this model (first SR phase, intermediate USR phase and final SR phase). Sharp transition between the SR and USR phases may cause quantum corrections like large one-loop corrections in the curvature power spectrum \cite{Cai:2018,Firouzjahi-1,Firouzjahi-2}. Moreover, the transition procedure could affect the size of local non-Gaussianity \cite{Cai:2018}. Studying of possible quantum corrections and non-Gaussianity in this framework are beyond the scope of the present work and we leave them for future works.

The background evolution through the concurrent solving of Eqs. (\ref{eq: FR-eqn2}) and (\ref{eq:KG-NC}) could be evaluated. Thence, Fig. \ref{fig:e1,e2,H} embodies plots of scalar field $\phi$ and its derivative $\phi_{,N}$, thereto the first $\varepsilon_1$ and second $\varepsilon_2$ slow- roll parameters against the $e$-fold number $N$. It is obvious from the mentioned plots that, in the USR region the velocity of the inflaton approaches to zero and $\varepsilon_1$ undergoes a decrease, which yields an increase in the curvature perturbations power spectrum (see Fig. \ref{fig:ps}). The slow-roll conditions ($\{\varepsilon_1,\varepsilon_2\}\ll1$) all over the USR region are obeyed by  $\varepsilon_1$, whereas broken by $\varepsilon_2$ momentarily. We also checked that the swampland criteria are satisfied in our model.

In addition, the curvature perturbations power spectra could be computed by way of solving the MS equation (\ref{eq:MS})
for Cases of the model (notice Table \ref{tab2} and Fig. \ref{fig:ps}). The portrayed spectra
of ${\cal P}_{\cal R}$ at CMB scale are confined in the observational constraint of Planck 2018
(${\cal P}^{*}_{\cal R}\simeq2.1 \times 10^{-9}$), while they enhance around seven order of magnitude in the USR region to produce PBHs.

Three Cases of PBHs mass spectra according to parameter Cases of the model have been attained
via the numerical solutions of MS equation. The consequent PBHs from the parameter Case A could be a promising claimant for $99\%$ of DM content. The anticipated PBHs for the parameter Case B in the permissible zone of the OGLE data \cite{OGLE-1,OGLE-2}, could be the origin of the ultrashort-timescale microlensing events. The foretold PBHs concerning the parameter Case C, in the sensitivity zone of LIGO-VIRGO  could be considered as the source of detected GWs by these detectors. The consequent values for PBHs abundance  $f_{\rm PBH}$ and mass $M_{\rm PBH}$ are arranged in Table \ref{tab2} and mapped in Fig. \ref{fig-fpbh}.

Thereafter, the spectra of $\Omega_{\rm GW_0}$ (the current density parameter of coeval GWs with PBHs) have been obtained  and plotted in Fig. \ref{fig-omega} for all Cases. The predicted $\Omega_{\rm GW_0}$ spectrum from parameter Case A, has lied in the sensitivity domain of LISA, whereas the resultant $\Omega_{\rm GW_0}$ spectra for Cases B and C have situated in the sensitivity domain of  SKA detector (see Fig. \ref{fig-omega}). Thus, the rectitude of this model could be verified in light of the approaching data of these detectors.

In the last stage, the power-law slopes of $\Omega_{\rm GW_0}$ spectra ($\Omega_{\rm GW_0} (f) \sim f^{n} $) \cite{fu:2020,Xu,Kuroyanagi} at different frequency regions have been estimated for all Cases of the model. Table \ref{table:GWs} embodies the peak frequencies of  $\Omega_{\rm GW_0}$ spectra and the estimated values for the power indexes $n$ for all Cases of the model in three frequency regions like $f\ll f_{c}$,  $f<f_{c}$ and  $f>f_{c}$.
It can be inferred that, the consequent quantities for $n$ in the infrared region $f\ll f_{c}$ could be compatible with the logarithmic equation $n=3-2/\ln(f_c/f)$  acquired in \cite{Yuan:2020,shipi:2020,Yuan:2023}.
%========================================Acknowledgements=====================================================
\section{Acknowledgements}\label{sec7}
The authors thank the referee for his/her valuable comments and they would like to thank Kazem Rezazadeh for helpful discussions as well. This work has been  supported financially by Vice President for Research and Technology, University of Kurdistan.

%===========================================Refrence======================================================

\end{document}